\documentclass[]{article}
\usepackage{amsfonts}
\usepackage{amssymb}
\usepackage{amsmath}
\usepackage{epsfig}
\usepackage{graphicx}

\def\Z{\mathbb Z}

\def\b{\bar}

\def\m{\mu}
\def\n{\nu}

\def\t{\tau}

\def\~{\widetilde}

\def\bY3{\bar Y_{,3}}
\def\Y3{Y_{,3}}
\def\z{\zeta}
\def\Z{{\b\zeta}}
\def\Y{{\bar Y}}

\def\`{\dot}
\def\be{\begin{equation}}
\def\ee{\end{equation}}
\def\bea{\begin{eqnarray}}
\def\eea{\end{eqnarray}}

\def\fn{\footnote}

\def\mn{{\mu\nu}}

\begin{document}
 \baselineskip=11pt
\title{Twistor String
Structure of the Kerr-Schild Geometry and Consistency of the
Dirac-Kerr System. \hspace{.25mm}\thanks{Paper dedicated to the
Jubilee of Professor S.D. Odintsov}}
\author{\bf{A. Burinskii}\hspace{.25mm}\thanks{\,e-mail address:
bur@Ibrae.ac.ru}
\\ \normalsize{Gravity Research Group, NSI, Russian Academy of Sciences}
\vspace{2mm}}
\date{}
\maketitle
\begin{abstract}
Kerr-Schild (KS) geometry of the rotating black-holes and spinning
particles is based on the associated with Kerr theorem twistor
structure which is defined by an analytic curve $F(Z)=0$ in
 the projective twistor space $Z\in CP^3 .$

On the other hand, there is a complex Newman representation which
describes the source of Kerr-Newman solution as a "particle"
propagating along a complex world-line $X(\t)\in CM^4,$ and this
world-line determines the parameters of the Kerr generating
function $F(Z).$ The complex world line is really a world-sheet,
 $\t=(t + i\sigma),$ and the Kerr source may be considered as a
complex Euclidean string extended in the imaginary time direction
$\sigma$. The Kerr twistor  structure turns out to be adjoined to
the Kerr complex string source, forming a natural twistor-string
construction similar to the Nair-Witten twistor-string.

We show that twistor polarization of the Kerr-Newman solution may
be matched with the {\it massless} solutions of the Dirac
equation, providing consistency of the Dirac-Kerr model of
spinning particle (electron). It allows us to extend the
Nair-Witten concept on the scattering of the gauge amplitudes in
twistor space to include massive KS particles.

\end{abstract}

\section{ Introduction}  Kerr-Schild (KS) geometry is a
background for the rotating black-hole (BH) solutions. On the
other hand, the Kerr-Newman solution has gyromagnetic ratio $g=2,$
as that of the Dirac electron, and the KS geometry acquires
central role as a model of spinning particle in gravity, in
particular, as a  Dirac-Kerr model of electron \cite{DirKer}.
Consistency of Quantum theory with Gravity is one of the principal
problems of modern physics, and in this paper we discuss a way to
progress in this direction, showing that consistency of the
Dirac-Kerr system for a massive particle (electron) may be
provided by a twistor-string source formed by the {\it massless}
solutions of the Dirac equation aligned with the twistor structure
of the KS background. In many respects this twistor-string is
similar to the Nair-Witten twistor-string model \cite{BurTwi},
which allows us to join to the Nair-Witten concept on the
scattering of the gauge amplitudes in twistor space
\cite{Nai,Wit}, extending this concept to include massive
particles.

 The KS solutions in 4D are algebraically special
solutions of type D which has two (doubled) Principal Null
Congruences (PNC) corresponding to geodesic lines of outgoing or
ingoing photons.
 Tangent vectors to these congruences, $k_\m(x)^\pm ,$ are null
 and determine two different coordinate system of the black-hole
solutions related to the `in' or `out' congruence, \cite{MTW}. KS
metrics have very simple KS form \be g_\mn =\eta_\mn - 2H k_\m
k_\n, \label{KS}\ee where $\eta_\mn $ is metric of auxiliary
Minkowski space-time, and the in (or out) null vector field
$k^\m(x), \ x\in M^4$ determines symmetry of space-time, in
particular, direction of gravitational `dragging`. For rotating BH
geometry the congruence is twisting which causes the difficulties
for its derivation and analysis.
The obtained exact solutions of the Einstein-Maxwell system of
equations \cite{DKS} indicate that electromagnetic (em) field is
not free with respect to the choice of the in or out
representation. The direction of congruence related with em field
must be the same as it is for metric, i.e. vector potential of em
field $A_\m$ has to satisfy the alignment condition , \cite{EM3p},
$A_\m k^\m =0.$ Since the classical em-fields are determined by the
retarded potentials, the in-out symmetry of gravity turns out to be broken
and the solutions of the Einstein-Maxwell system have to be based
on the out-congruence, which has important physical consequences.
\fn{s may mistakenly be associated  with white BH. The necessity of choice of
out-congruence is also argued in \cite{SHW} from quantum point of view.}
 The structure of Kerr congruence for the stationary Kerr-Newman
 solution at rest is shown in Fig.1.
\begin{figure}[ht]
\centerline{\epsfig{figure=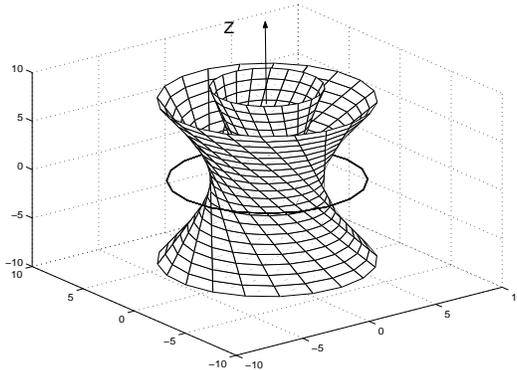,height=5cm,width=7cm}}
\caption{The Kerr singular ring and the Kerr congruence. }
\end{figure}
It displays very specific twosheetedness of the KS space-time. The
Kerr singular ring represents a branch line of space-time, showing
that congruence is propagating from negative sheet of metric onto
positive one, and  the outgoing congruence is analytic extension
of the ingoing one.
The KS metrics have a rigid connection with metric of auxiliary
Minkowski space-time $\eta_\mn $ and are practically unfastened
from the position of horizon. KS approach allows one to obtain
solutions which are form-invariant with respect to the position of
horizons, and even to its presence, which allows one to consider
the black-holes and spinning particles without horizon on the
common grounds \cite{RenGra}, as well as perform the analysis of
deformations of the black-hole horizon by electromagnetic
excitations \cite{EM3p,BEHM1}. Besides, the twosheetedness of the
Kerr space reproduces the properties of BH  space-times which are
desirable from quantum point of view \cite{SHW}.
\section{The Kerr theorem} Kerr congruence forms a fiber bundle of
the KS space-time which is determined by the Kerr theorem
\cite{BurKer,BurNst,Pen,KraSte}. The lightlike fibers of the Kerr
congruence are {\it real} twistors (intersections of the complex
conjugated null planes \cite{BurNst}). The Kerr theorem gives a
rule to generate geodesic and shear-free (GSF) null congruences in
Minkowski space-time $M^4.$ Due to the specifical form of the KS
metrics, these congruences turn out to be also geodesic and
shear-free in the curved KS space-times which justifies
application of the KS formalism \cite{DKS} to solutions of the
Einstein-Maxwell field equations.
 The Kerr theorem states that any GSF congruence in $M^4$
 is determined by some holomorphic generating function $F(Z)$ of
the projective twistor coordinates \be Z =( Y, \quad \z - Y v,
\quad u + Y \Z \ ) \label{PTw} ,\ee being an analytic  solution of
 the equation $F(Z)=0$ in projective twistor space, $Z\in CP^3 .$
\fn{Twistor coordinates are defined via the null Cartesian
coordinates $ 2^{1\over2}\z = x + i y ,\quad 2^{1\over2} \Z = x -
i y , \quad 2^{1\over2}u = z - t ,\quad 2^{1\over2}v = z + t .$}
 The variable $Y$ plays especial role, being
projective spinor coordinate $ Y= \pi^2/\pi^1 $ and simultaneously
the projective angular coordinate \be Y=e^{i\phi}\tan \frac \theta
2 \label{Y0} .\ee The dependence $Y(x)$ is the output of the Kerr
theorem which determines the Kerr congruence as a field of null
directions \be k_\m dx^\m = P^{-1}(du +\Y d\z + Y d\Z - Y \Y dv).
\label{kpm}\ee The null field $k^\m$  may also be expressed in
spinor form \be k^\m =\bar \pi \sigma^\m \pi \label{km} \ee via
the Pauli matrices $\sigma^\m ,$ or in the terms of spinor
components \cite{Pen,Wit,Nai} \be k_{a\dot a}= \sigma^\m_{a\dot
a}k_\m = \pi _a \bar\pi _{\dot a}. \ee Each real null ray
represents a twistor which is fixed by projective twistor
coordinates (\ref{PTw}), or by homogenous coordinates \be Z^\alpha
= \{ \pi^a, \mu _{\dot a}.\} \label{Tw} \ee
 The spinor $\pi^a$ determines the null direction $k^\m$,
and spinor $\mu _{\dot a}=x_\n \sigma^\n_{\dot a a} \pi^a$ fixes
the position (equation) of a real null ray (or of the
corresponding complex null plane).
 {\bf Quadratic in  $Y$ functions
 $F(Y)$} determine two roots $Y^\pm(x)$ of the equation $F=0$
 which corresponds to twosheetedness of the Kerr space-time.
It was shown in seminal work \cite{DKS} that any quadratic
generating function $F(Y)$ of the Kerr theorem determines a class
of the exact solutions of the Einstein-Maxwell field equations.
The case of quadratic in $Y$ functions $F(Y)=A(x)Y^2 +B(x)Y +C(x)
\equiv A(x)(Y-Y^+)(Y-Y^-)$ was studied in details
\cite{BurNst,BurMag}. In fact, it describes the Kerr-Newman
solution in a general position with arbitrary spin-orientation and
Lorentz boost. This information is encoded in the coefficients
$A,B,C ,$ and the KS formalism allows one to write down the
corresponding form of congruence and the exact solutions for
metric and em field.
\section{Complex string as source of Kerr geometry} Structure of
the Kerr-Newman solution admits a complex interpretation
(suggested by Newman \cite{New}) as being generated by a source
propagating along a complex world-line $X^\m(\t) \in CM^4,$  and
parameters of the quadratic in $Y$ function $F(Y)$ may be easily
determined from parameters of the world-line \cite{BurNst,BurMag}.
Since a complex world-line $X^\m(\t)$ in $CM^4$ is parametrized by
complex time $\t=t +i\sigma ,$ it is really a complex world-sheet
with target space $CM^4 ,$ \cite{BurStr}. Thereby, the complex
Kerr-Newman source is equivalent to some complex string. The KS
twistor structure may be described by a complex retarded-time
construction, in which twistors of the Kerr congruence are the
real sections of the complex light cones emanated from the complex
world-line\fn{Newman suggested this construction for $CM^4 ,$
\cite{New}, while the application to curved space-times demands
the KS representation for metric \cite{BurNst,BurTwi}.}. In this
way the Kerr-Newman solution may be generalized to solutions of a
broken $N=2$ supergravity \cite{BurSup}. Projection of the {\it
real} KS twistors onto complex world line selects on the
world-sheet a strip $Im \ \t \in [-a,a],$ forming an open
Euclidean string extended along the complex time direction
$\sigma=Im \t$. The parameter $\sigma$ is linked with angular
directions of  twistor lines, $\sigma = a\cos \theta .$\fn{In
fact, each point of the complex string has adjoined complex
twistor parameter $Y $ with the additional freedom of the
rotations $e^{i\phi}$ around z-axis. Up to this $U(1)$ symmetry,
each point of the complex Kerr string has an adjoined real twistor
line of the Kerr congruence, forming the Kerr twistor-string
structure \cite{BurTwi}.} Condition of the existence of real slice
for the complex twistors determines the end points of the string
$\t = t \pm ia $ \cite{BurStr}. In the same time, consistency of
boundary conditions demands orientifolding this string, by
doubling of the world-sheet and turning it into a closed but
folded string \cite{BurStr,BurTwi}. The joined to the end points
of this string twistors form two especial null rays emanated in
the North and South directions, $k_N=\bar \pi_N \sigma^\m \pi_N $
and $k_S=\bar \pi_S \sigma^\m \pi_S . $ \cite{DirKer}. They play
peculiar role in the KS geometry, controlling parameters of the
function $F ,$ and therefore, the twistorial structure in whole.
The corresponding end points of complex string may be marked by
quark indices $\Psi^\alpha_{N(S)} . $ It was obtained in
\cite{DirKer,BurTwi} that the null directions $k_N$ and $k_S$ may
be set in the one-to-one correspondence with solutions of the
Dirac equation $\Psi=(
 \phi _\alpha,
\chi ^{\dot \alpha}).$ Putting $\pi_N=\phi$ and $ \pi_S = \bar
\chi,$ one sees that the Dirac wave function $\Psi$  manages the
position, orientation and boost of the Kerr source. It gives a
combined Dirac-Kerr model of spinning particle, in which  twistor
structure is controlled by the solutions of the Dirac equation
\cite{DirKer}. Plane waves of the Dirac wave function propagate
along the null directions $k_N$ and $k_S.$
The essential drawback of this model is that the Dirac equation
and its wave functions are considered in $M^4$, instead of the
consistent treatment  on the KS background. However, the exact
solutions of the massive Dirac equation on the Kerr background are
unknown, and moreover, there are evidences that they cannot be
consistent with KS structure in principle. Note also that the
plane waves are inconsistent with the Kerr background and there is
a related problem with consistency of the conventional Fourier
transform on the twosheeted Kerr background.
\section{Consistency of the Dirac-Kerr system}
Consistent solutions of the Dirac equation may easily be obtained
for the {\it massless} fields which are aligned with the Kerr
congruence. For the aligned to PNC solutions, the Dirac spinor
$\Psi_D ^\dag = (\phi,\chi) =( A, \ B, \ C, \ D )$ has to satisfy
the relation $\Psi_D k_\m \gamma^\m =0 ,$ which yields,
\cite{BurSup,EinFink}, $A=D=0,$ and the functions $B$ and $C$ take
the form (\cite{BurSup}, App. B)
\begin{equation}
B=  f_B (\bar Y, \bar \tau)/ \bar{\tilde r},\qquad C= f_C ( Y,
\tau)/\tilde r,\quad \tilde r=r+ia\cos \theta,
\label{BC}\end{equation} where $f_B$ and $f_C$ are arbitrary
analytic functions of the complex angular variable $Y$ and the
retarded-time $\tau = t - \tilde r ,$  obeying the relations
$\tau,_2=\tau,_4 =0, $ and $Y,_2=Y,_4=0$. Due to this analyticity,
the wave solutions have poles at some values of $Y,$ or a series
of such poles $f_B= \sum_i a_i /(Y-Y_i), $ leading to singular
beams along some of twistor lines \fn{It is valid for em-field
too, \cite{BEHM1}}. It turns out that the treatment of the
massless solutions resolves four problems:
1) existence of the exact self-consistent solutions,
2) the Dirac plane waves on the KS background turn into the
lightlike beams concentrating near singular twistor lines with
directions $k_N$ and $k_S.$
3) there appears natural relation to the massless core of
superstring theory \cite{GSW} and to the massless Nair-Witten
twistor models for scattering, \cite{Wit,Nai}.
4) the Dirac massless solutions form the currents and lightlike momenta $p(Y)$
distributed over sphere, $Y\in S^2 , $ and the total nonzero mass
appears after averaging over sphere.
The real source of the Kerr-Newman solution is a disk, D2-brane boundary (base)
of a ``holographic" twistor bundle. The
total mass has a few contributions \cite{RenGra}, including the
source itself and the mass-energy of the fields distributed  over
null rays. It depends on $Y \in S^2$ which may be factorized into
angular part $e^{i\phi},$ and parameter $\sigma=a\cos \theta=
a\frac {1-Y\Y}{1+Y\Y}$ along the complex Kerr string.
Orientifolding is expected to be equivalent to normal ordering \be
p^\m=<:\bar\Psi \gamma^\m\Psi:
> = \int _{S^2} \m(Y) d Y d\Y :\bar \Psi
\gamma^\m\Psi : .\ee  In the rest frame \be m = <p^0(Y)>=
\int_{S^2}\m(Y)dYd\Y p^0(Y). \ee
  Therefore, mechanism of
the origin of mass from the massless solutions is similar to that
of the dual (super)string models and corresponds to an initial
Wheeler's idea of  a geon \cite{Bur0} (``mass without mass''
obtained from the averaged em field + gravity), as well as to the
initial Ramond idea on averaging of the local string structure in
the dual string models \cite{Ram}. The corresponding wave
solutions on the Kerr background have singular beams which take
asymptotically the form of the exact singular pp-wave solutions by
A. Peres \cite{BurAxi}.
\section{Solutions with singular Beams as analogs of the plane and spherical waves}
Note, that in general the conventional electromagnetic (em) plane
waves are not consistent solutions in a curved space-time. The
closest gravitational analogs of the em plane waves are the so
called `` exact plane wave'' solutions which are singular at
infinity \cite{HorSte,Coley}. More general are the plane-fronted
wave solutions (pp-waves) which have the unique symmetry along a
covariantly constant null direction $k^\m $ (one can set $k^\m=du,
\ u=z-t$ and, therefore, belong to the KS class $g_\mn = \eta_\mn
+ H(u) k_\m k_\n.$ They may be regular at infinity, but should
have one or more poles in the finite region of the orthogonal to
$k^\m$ plane $x+iy.$ These poles determine the position of
singular beams. Finally, there is a third type of singularity at
the front of solution, at a fixed value of the retarded time $u ,$
\cite{HorSte,Coley}. Similar, the conventional spherical (or even
the ellipsoidal) harmonics cannot be exact em solution for the
curved Kerr background. So, in principle, the conventional Fourier
analysis cannot be used on the KS space-times. The KS solutions
with singular em beams aligned with a constant null vector may be
exact solutions to the coupled system (em + gravity) and be
considered as analogs of the exact plane waves. Similar, the
solutions with singular beams aligned with one or many rays of the
Kerr congruence turn out to be analogs  of the conventional smooth
harmonics on the KS space-time. They form overfilled system of
coherent states, and appear in many problems related with the
exact solutions on the KS background \cite{BurSup,BEHM,BEHM1}.
Such beams lead to some unexpected physical consequences. In
particular, the beam-like solutions have very strong back reaction
on metric and break the topological structure of the black-hole
horizons, leading to its topological instability  with respect to
electromagnetic excitations \cite{BEHM,BEHM1}. The KS approach
gives as a broad class of the exact em solutions with beams which
are consistent with the KS gravity and aligned with twistor
structure.

\section{Multi-particle KS solutions and Quantum Gravity}

There is a very broad class of exact em solutions with beams
coupled with KS gravity. They are based on the Kerr theorem
related with the Kerr generating functions $F(Y)$ of higher
degrees in $Y ,$ \cite{Multiks}. The corresponding KS congruences
form multisheeted Riemann surfaces, and the resulting space-times
turn out to be multisheeted too, which generalizes the known
twosheetedness of the Kerr geometry.

If we have a system of $k$ particles with known parameters $q_i$,
one can form the function $F$ as a product of the $k$ given blocks
$F_i(Y)=F(Y|q_i)$ with a known dependence $F(Y|q_i) ,$
\be F(Y)\equiv \ \prod _{i=1}^k F_i (Y) \label{multi}. \ee
The solution of the equation $F=0$ acquires $2k$ roots $Y_i^\pm$,
leading to 2k-sheeted twistor space.

\begin{figure}[ht]
\centerline{\epsfig{figure=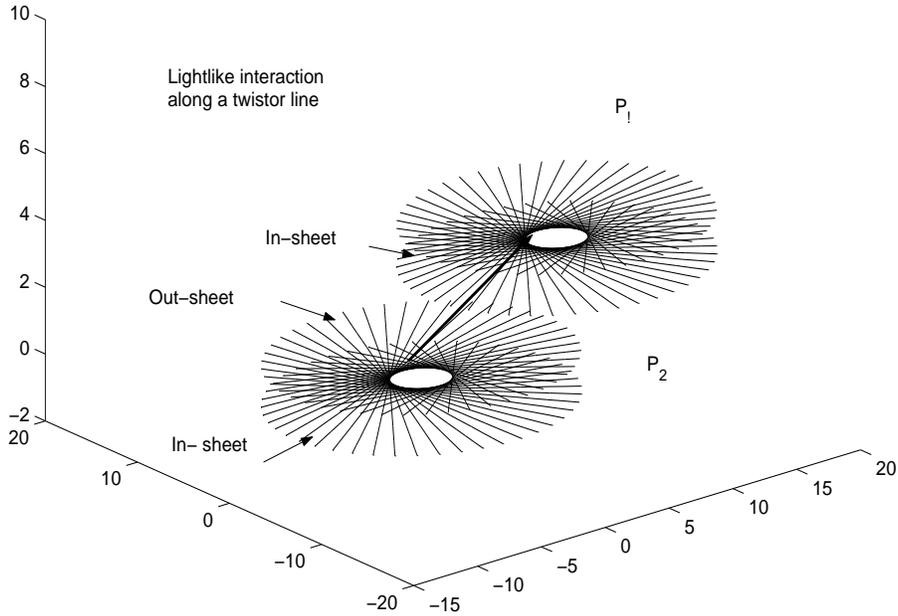,height=10cm,width=12cm}}
\caption{The lightlike interaction in Multi-sheeted twistor space
 via a common twistor line connecting the out-sheet of one particle
 to the in-sheet of another.}
\end{figure}

The twistorial structure on the i-th $(+)$ or $(-)$ sheet is
determined by the equation $F_i=0$ and does not depend on the
other functions $F_j , \quad j\ne i$. Therefore, the particle $i$
does not feel the twistorial structures of other particles.
Similar, singular sources of the $k$ Kerr's spinning particles are
determined by equations $F=0, \ d_Y F=0$ which acquires the form
\be \prod _{l=1 }^k F_l =0, \qquad   \sum ^k_{i=1} \prod _{l\ne
i}^k F_l d_Y F_i =0 \label{leib} \ee
 and splits into k independent relations
\be F_i=0,\quad \prod _{l\ne i}^k F_l d_Y F_i =0 \label{kind}, \ee
showing  that i-th particle does not feel also the singular
sources of other particles.  The  space-time splits on the
independent twistorial sheets, and therefore, the twistor
structure related to the i-th particle plays the role of  its
``internal space''.
It looks wonderful. However, it is a natural generalization of the
well known twosheetedness of the Kerr space-time which remains one
of the mysteries of the Kerr solution for the very long time.
In spite of independence of the twistor structures positioned on
the different sheets, there is an interaction between them via
 the singular lightlike beams (pp-strings) which appear on the
 common twistor lines connecting the different sheets of the particles
\cite{Multiks} and play the role of propagators in the Nair-Witten
concept on scattering amplitudes in twistor space \cite{Wit}.
Multisheetedness of the KS spacetimes cannot be interpreted in the
frame of classical gravity, however it has far-reaching
consequences for quantum gravity \cite{SHW}, in which metric is
considered as operator $\hat g _\mn $ leading to an ``effective
geometry'', \be <out|\hat g _\mn |in> = g _\mn <out|in> .\ee If we
consider the Kerr-Newman solution for an isolated BH or spinning
particle, we obtain the smooth space-time apart from the Kerr
singular ring. However, the exact solution for the Kerr-Newman
source surrounded by remote massive or massless particles will
contain series of singular beams along the twistor lines of Kerr
congruence connecting the Kerr source with these remote particles.
Analysis of the action of such beams on the BH horizon, given in
\cite{BEHM1}, showed that the beams have very strong back reaction
on metric and even the very weak beams pierce the horizon changing
its topology. The beams, caused by external particles
\cite{Multiks}, as well as the beams caused by the em zero point
field,\cite{EM3p,BEHM}, lead to a fine-grained topological
fluctuations of horizon. The exact wave solutions on the KS
background show that the beams are coherent and freely overpass
the BH horizon \cite{BurPass}. It revokes the loss of information
inside of BH.

Considering metric as an operator, one has to give an operator
meaning to the Kerr theorem. Immediate way is to consider the complex
position of the Kerr source $X^\m(\t)$ and its velocity $\dot X^\m(\t)$
as coordinates of a quantum oscillator in spirit of the original Ramond
paper \cite{Ram}.  It leads to the operator meanings of the
the equation $F=0$ via coefficients of the function $F(Y),$ and suggests
an operator approach to the Kerr-Schild equations \cite{DKS}.

\section{KS twistor structure and the
Nair twistor WZW model}  Nair considers $S^2$ as a momentum space
of the massless gauge bosons and constructs a classical WZW model
over the sphere $S^2$ which is parametrized by the spinor
coordinates $\pi^a ,$ \cite{Nai}. This construction is close to
our field of null directions $k^\m (\pi^a)$ in $M^4 , $ which is
parametrized by $Y\in S^2, \ Y=\pi^2/\pi^1 .$ The suggested by
Nair and Witten treatment of the scattering amplitudes in twistor
space $CP^3$ assumes that the time evolution occurs in $M^4$ along
twistor null lines $Z=const.,$ while $S^2$ has only topological
duty, playing the role of 2D boundary (base) of the
``holographic'' twistor bundle of $M^4 .$ In the KS twistor
structure this boundary is the real disk-like source of the Kerr
solution, forming the area for complex fields on $S^2$ and the
currents. Introducing two-dimensional spinor fields $\Psi(\pi) $
on the sphere, Nair considers correlation function \be
<\Psi_r(\pi) \Psi_s(\pi ')>= \frac {\delta_{rs}} {\pi \pi'} \sim
\frac {\delta_{rs}} {Y-Y'}\label{corr}\ee and the local currents
$J^a=\bar\Psi t^a \Psi ,$ showing that OPE takes in homogenous
spinor coordinates the usual form of the corresponding Kac-Moody
algebra of the WZW model\cite{KniZam} with central extension $ k=
1.$ In this case the complex-analytic structure of WZW model is
lifted from $S^2$ to twistor space $CP^3$ over $M^4.$ Twistor
string structure of the KS geometry has much in common with the
Nair WZW model and Witten's twistor string B-model, as well as
many specifical features. However, in all the cases twistor
approach displays the principal advantage, providing a natural
extension of CFT  from $S^2$ to $M^4.$  Since the massless KS
twistor space allows one to describe  massive spinning particles,
the Nair-Witten treatment of the scattering amplitudes of the
gauge fields may be extended in the KS geometry to describe
scattering of the gauge fields on the massive black-holes and
spinning particles.
\section{Conclusion}
 We have showed that consistency of the Dirac-Kerr model for a {\it massive}
 spinning particle in gravity (in particular electron) has to be based on
the {\it massless} solutions of the  Dirac equation aligned with
the KS twistor background. It provides a natural twistor string
structure of the KS geometry and generalization of the Nair-Witten
approach to scattering of the gauge amplitudes in twistor space to
include massive particles formed from the KS twistor-string
bundle.

The considered here aligned solutions of the
 massless Dirac equation were extracted from a fermionic part of the
 super-Kerr-Newman solutions to broken $N=2$ supergravity \cite{BurSup}, which
 rises interests in the corresponding solutions to supergravity. Self-consistency
 of the considered aligned solutions shows that the KS twistor structure gives a
 clue to unification of quantum theory and gravity.

\end{document}